\documentclass{aa}
\usepackage{graphicx}
\begin{document}
   \title{Propagation of non-linear circularly polarised Alfv\'{e}n waves
    in a homogeneous medium}


   \author{Rim Turkmani \thanks{{\em Present address: Space and Atmospheric 
Physics, The Blackett Laboratory, Imperial College, London, England. } }
          \and
          Ulf Torkelsson
          }


   \institute{Chalmers University of Technology/G\"oteborg University, 
   		Department of Astronomy \& Astrophysics, \\
		S-412 96 Gothenburg, Sweden
             }
   \date{Accepted 23-07-03}
   \authorrunning{Turkmani \& Torkelsson}
   \titlerunning{Circularly polarised Alfv\'{e}n waves
    in a homogeneous medium}

   \abstract{
   We study the evolution of non-linear circularly polarised Alfv\'{e}n waves by 
   solving numerically the time-dependent equations of magnetohydrodynamics (MHD) 
   in one dimension. 
   We examine the behaviour of the waves and find that different physical 
   mechanisms are relevant in different ranges of $\beta$.
   In a low $\beta$ plasma the wave may undergo a parametric decay.  This is because
   the wave excites a density enhancement that travels slower than the wave
   itself and thus interacts with the wave.  
   When $\beta \ge 1$ the density enhancement does not interact with the wave and no 
   decay takes place, instead the Alfv\'en wave is reflected against the density 
   enhancement.
   The reflection zone propagates with the speed $\frac{1}{n} \, v_A$.
   Because of that the magnetic flux is conserved which results in an amplification
   of the oscillating magnetic field by a factor $\frac{1}{n}$. We find that
   $n$ depends on $\beta$, and that in particular it is $ \le \, 1$ 
   for values of $\beta \sim 1$ and  $ \ge \, 1$ for $\beta \gg 1$.
   We discuss the relevance of these mechanisms to the acceleration of the 
   solar wind, and the triggering of MHD turbulence in the polar wind region.
   In particular these simulations can explain the presence 
   of inward propagating Alfv\'en waves in the solar corona. 
   }

   \maketitle
\section{Introduction}

The expanding coronal model proposed by Parker (\cite{Parker}) predicted the 
existence of the solar wind.
Observations by space probes showed later that there are two forms of 
solar wind: the slow solar wind (up to 400 km/s) which fits the Parker 
model, and the fast solar wind (up to 800 km/s) which emanates from 
the coronal holes, regions in the solar corona with open magnetic field 
lines and low density.

Large amplitude, low frequency, Alfv\'{e}n waves have been observed in the 
solar corona for over 30 years (e.g. Belcher \& Davis \cite{BD}).
During the last decade Ulysses has provided plasma and magnetic field 
measurements that have allowed extensive investigations of the behaviour 
of Alfv\'{e}nic turbulence in the high-latitude solar wind.  The data shows a 
strong correlation between the fluctuations in velocity and magnetic fields
(Smith et al. \cite{smith}), and revealed the presence of both  inward and outward 
directed Alfv\'{e}n waves (e.g Bavassano et al. \cite{BV}). While the outward going waves are 
expected to be generated at the coronal base, the source of the 
inward going waves in the solar wind is not yet understood.

Alfv\'{e}n waves  play a crucial role in several 
models for the acceleration of the fast solar wind
(e.g. Leer et al. \cite{LHF}).
An attractive feature of the Alfv\'{e}n waves is that they can propagate 
over vast distances since they are incompressible to lowest order, and 
therefore do not dissipate easily.
However, dissipative damping is required at some point to avoid too high 
wind velocities (e.g. Holzer et al. \cite{HL}).

In two dimensions phase mixing
due to a transverse gradient in the phase velocity (e.g. Heyvaerts \& 
Priest \cite{HP}) can lead to a strong damping of the wave.  
This mechanism is not available in one dimension though, and one has then 
rather to consider the nonlinear coupling of the Alfv\'en wave with other 
modes (e.g. Wentzel \cite{V}).
A close examination of the properties of Alfv\'{e}n waves shows that a 
linearly polarised  Alfv\'{e}n wave is compressible to second order though, 
because the magnetic pressure, $B_{\perp}^2 / 2 \mu_0$, is modulated on half 
the wave length of the Alfv\'{e}n wave itself 
(Alfv\'{e}n \& F\"{a}lthammar \cite{AF}). 
This opens up the possibility that a high-amplitude Alfv\'en wave can
steepen, which was demonstrated analytically by Cohen \& Kulsrud (\cite{CK}).  
As the wave steepens, it forms current sheets at the nodes of the 
fluctuating magnetic field.
This effect has  been studied in numerical simulations by Boynton 
\& Torkelsson (\cite{BTT}) and Ofman \& Davila (\cite{ofman}).

In a circularly polarised Alfv\'{e}n wave, on the other hand, the magnetic 
pressure is constant along the wave, which is the physical reason why it 
is an exact solution of the nonlinear MHD equations. 
However, the wave can still decay via a parametric instability (e.g. Sagdeev
\& Galeev \cite{sg}), 
which is usually less important than the wave steepening for a 
linearly polarised Alfv\'en wave. 
In this instability a forward propagating Alfv\'{e}n wave in the presence
of a density fluctuation generates a forward propagating acoustic wave and
a backward propagating Alfv\'en wave.

Parametric decay has many applications in plasma physics and astrophysics. 
In coronal physics it has been proposed to be a possible mechanism to trigger MHD 
turbulence in regions with relatively smooth density profiles like the polar wind 
region and to account for the small compressible fluctuations in the solar wind 
(Goldstein \cite{gold}). 
It is also a candidate for generating the inward propagating Alfv\'{e}n waves
in the solar wind (e.g. Tu \& Marsch \cite{tu})

\begin{figure}
\includegraphics[width=8.8cm]{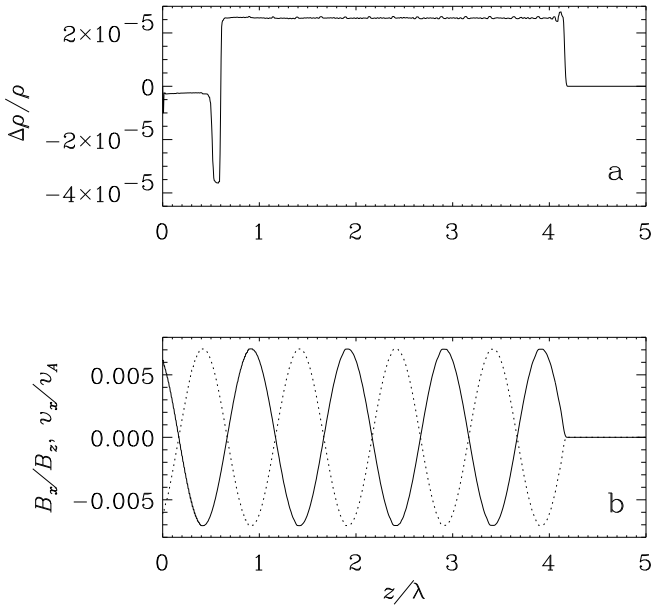}
\caption{Low amplitude magnetohydrodynamic waves propagating through a homogeneous
medium of low $\beta$ (Run 1a).
{\bf a} $ \Delta \rho /\rho= \frac{\rho - \rho_0}{\rho_0}$ 
versus $z/\lambda$ at $t/P = 4.2 $. 
{\bf b} $B_x/B_z$ (solid line) and $v_x/v_{\rm A}$ (dotted line) versus
$z/\lambda$ at the same time.}
\label{ph3a}
\end{figure}

The parametric decay of circularly polarised Alfv\'en waves has been studied 
both analytically (e.g. Cohen \cite{C}, Derby \cite{derby}, Goldstein \cite{gold}, 
Jayanti \& Hollweg \cite{jay}) and numerically by several groups 
(e.g. Del Zanna et al. \cite{zanna}, Malara et al. \cite{mal}, Pruneti \& Velli \cite{Prun},
Ghosh \& Goldstein \cite{ghol}, Ghosh et al. \cite{Ghoh}, Umeki \& Terasawa \cite{U}).
Except for the work by Pruneti \& Velli (\cite{Prun}) these studies have been 
restricted to Alfv\'en waves in homogeneous media.  Most of these
numerical simulations have studied how Alfv\'en waves of different
frequencies interact and generate turbulence, and determine the properties of
fully developed turbulence.
For these purposes it is appropriate to use a model with
periodic boundary conditions and introduce Alfv\'en waves that extend over
the entire grid through the initial state (Malara et al. \cite{mal} and Del
Zanna et al. \cite{zanna}).  

On the contrary we will study how a {\em fresh}
Alfv\'en wave falls prey to the parametric instability, which gradually
converts it into a backward propagating Alfv\'en wave and an acoustic wave. 
To do this we drive the Alfv\'en
wave on one of the boundaries of our model.  
Since there is no Alfv\'en wave in the interior
initially we are able to study how processes
at the propagating wave front eventually lead to the breakdown of the 
entire Alfv\'en 
wave further upstream, a phenomenon that cannot be found using periodic
boundary conditions.
Our approach is similar to that of Boynton \& Torkelsson (\cite{BTT}), 
who studied linearly polarised
Alfv\'en waves.
Since the processes that we study occur gradually it is important to use a 
grid that extends over many wave lengths to capture them, which limits us
to using a one-dimensional model.  The need for an very extended grid 
becomes clear if one compares the works by Torkelsson \& Boynton (\cite{TB})
and Ofman \& Davila (\cite{ofman}).  Torkelsson \& Boynton found a 
much stronger damping in their one-dimensional model than Ofman \& Davila did
in their two-dimensional model, but most of the damping occurred beyond 
the outer boundary of the Ofman \& Davila model.

A highly simplified numerical model such as the one that we present in this
paper cannot provide a realistic representation of the physics of Alfv\'en
waves in the solar wind, however it can still be useful since it allows us
to study a limited number of physical processes in detail.  The understanding
that we gain from this can then serve as a guide in interpreting some
aspects of more 
complex numerical simulations, such as the ones carried out by Tsiklauri et al.
(\cite{TNA}) and Tsiklauri \& Nakariakov (\cite{TN})
and Laveder et al. (\cite{lavedera}, \cite{lavederb}).

The plan of the paper is the following: In Sec. 2 we describe the basic MHD equations
and review the properties of the parametric decay. 
Our models and results are presented in Sect. 3.  We discuss how 
the results relate to the dynamics of Alfv\'en waves in the solar wind in Sect.
4 and we summarise our conclusions in Sect. 5.


\begin{figure*}
\centering
\includegraphics[width=17cm]{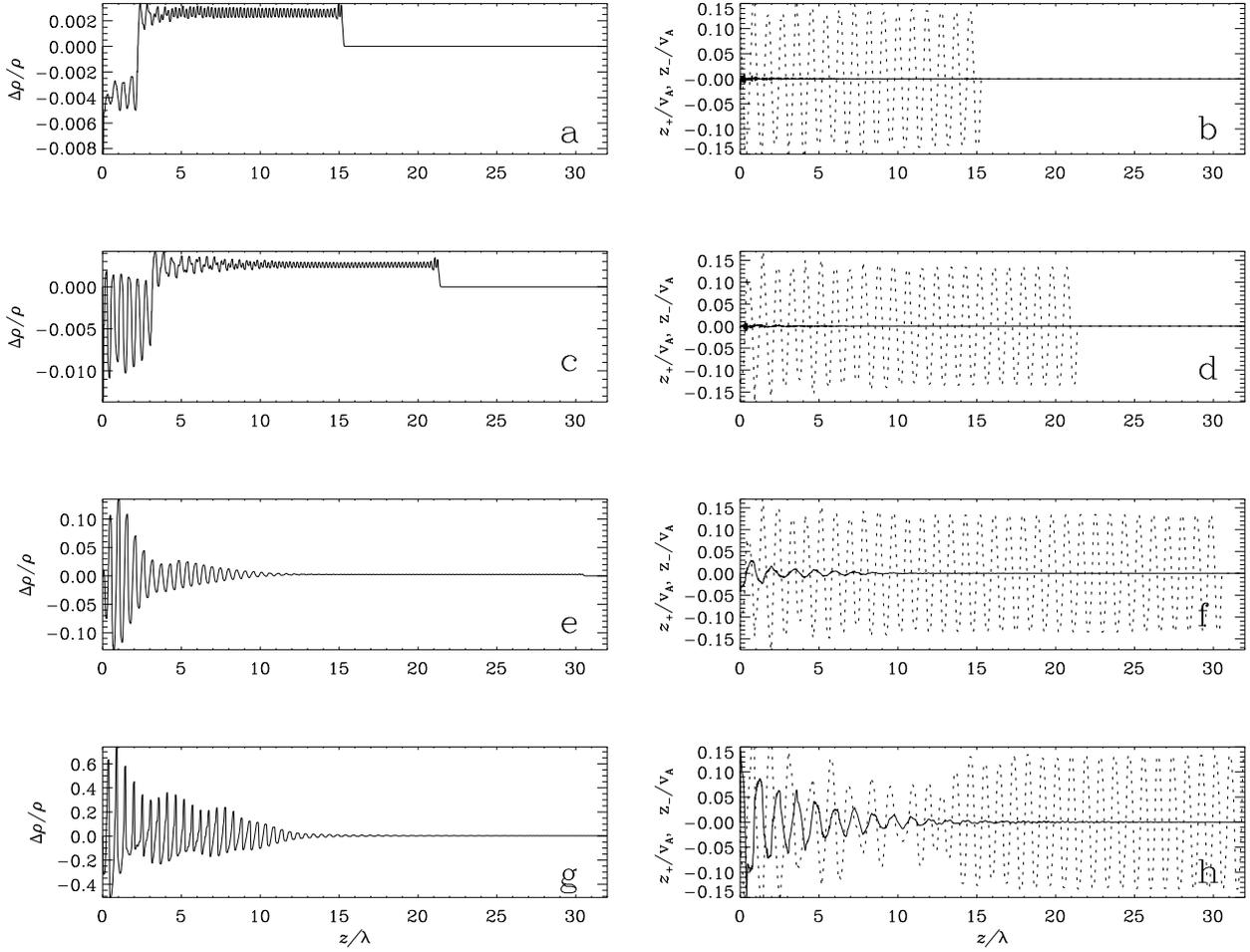}
\caption{Magnetohydrodynamic waves propagating through a homogeneous 
medium of low $\beta$ and $\eta = 0.07$ (Run 1b).
{\bf a, c, e, g} $ \Delta \rho / \rho=\frac{ \rho - \rho_0}{\rho_0}$
versus $z/\lambda$ at $t/P= 16.7, \, 23.3, \, 33.3, \, 46.7$.  
Note the changes in the $\Delta \rho/\rho$-scale between the frames.
{\bf b, d, f, h} $z_-/v_{\rm A}$ (solid line) and $z_+/v_{\rm A}$ (dotted line) 
versus $z/\lambda$ at the same times.}
\label{1b}
\end{figure*}


\section{Mathematical formulation}
\subsection{Fundamental properties of Alfv\'en waves}

The equations of ideal isothermal MHD in a homogeneous medium can be
written as

\begin{eqnarray}
\frac{\partial \rho}{\partial t} + \nabla \cdot (\rho {\bf v})=0 ,
\label{cont}
\end{eqnarray}

\begin{eqnarray}
\frac{\partial (\rho {\bf  v})}{\partial t}+\nabla \cdot ({\bf  v} \rho
{\bf  v})= -\nabla p + { \bf J} \times {\bf B},
\label{mom}
\end{eqnarray}

\begin{eqnarray}
\frac{\partial {\bf B}}{\partial t}=\nabla \times ({\bf  v} \times {\bf B}),
\label{induct}
\end{eqnarray}

\begin{eqnarray}
\nabla \cdot {\bf B}=0,
\end{eqnarray}

\noindent where $\rho$ is the density, ${\bf v}$ the velocity, $p$ the
pressure, $\bf B$ the magnetic field, and $\bf J = \nabla \times {\bf B}/ \mu_0$ 
the current density.
Eqs. (\ref{cont}) and (\ref{mom}) express the conservation of mass and momentum, respectively, 
and Eq. (\ref{induct}) is the induction equation. 
The constraint \begin{math} \nabla \cdot {\bf B}=0 \end{math} is fulfilled by 
Eq. (\ref{induct}) if it is imposed as an initial condition.
For fully ionised hydrogen we can write the equation of state as
\begin{equation}
  p = \frac{2\rho k_{\rm B} T}{m_{\rm H}},
\end{equation}
where $k_{\rm B}$ is Boltzmann's constant, $T$ the temperature and 
$m_{\rm H}$ the mass of the hydrogen atom.

In a homogeneous medium with a density $\rho_0$ and a background magnetic field 
${\bf B}=B_z {\bf \hat{z}}$ a circularly polarised forward propagating 
Alfv\'en wave is described by the transverse magnetic field
\begin{eqnarray}
{\bf B_{\perp}}=B_{\perp} [ \cos(kz -\omega t) {\bf \hat{x}}+ \sin(kz-\omega t) 
{\bf \hat{y}}],
\end{eqnarray}
and the velocity
\begin{equation}
  {\bf v_\perp} = - \frac{{\bf B_\perp}}{\sqrt{\mu_0 \rho_0}},
\label{v-B-rel}
\end{equation}
while for a backward  propagating Alfv\'en wave the velocity
is given by
\begin{equation}
  {\bf v_\perp} = \frac{{\bf B_\perp}}{\sqrt{\mu_0 \rho_0}}.
\end{equation}
In both cases the Alfv\'en wave obeys the dispersion relation
\begin{equation}
  \omega = v_{\rm A} k
\end{equation}
with the Alfv\'en velocity
\begin{equation}
v_A= \frac{ B_z}{\sqrt{\mu_0 \rho_0}}. 
\end{equation}
In order to separate forward and backward propagating 
Alfv\'en waves it is useful to introduce the Els\"asser variables

\begin{eqnarray}
{Z}_{\pm}=\left|{\bf v}_\perp \mp \frac{{\bf B}_\perp}{\sqrt{\mu_0 \rho}}\right|,
\end{eqnarray}
which describe the forward and backward propagating Alfv\'en waves,
respectively.  For a pure forward propagating Alfv\'en wave
$z_-$ does not exist while $z_+$ has twice the value of $v_\perp$
of the original wave.  
For the circularly polarised wave the magnetic pressure ${B_\perp}^{2}/(2 \mu_0)$ is 
the same everywhere inside the wave. 
This is the physical reason why an infinitely extended circularly polarised 
Alfv\'{e}n wave is incompressible and an exact solution of the nonlinear MHD 
equations, whereas the linearly polarised wave
\begin{eqnarray}
{\bf B_{\perp}}=B_{\perp} [\cos(kz -\omega t) {\bf \hat{x}}+ \cos(kz-\omega t)
{\bf \hat{y}}],
\end{eqnarray}
is compressible to second order, and therefore not an exact
solution of the nonlinear 
MHD equations.

In the one-dimensional problem that we study, there is only one additional 
wave mode, an acoustic wave obeying the dispersion relation
\begin{equation}
  \omega = c_{\rm s} k,
\end{equation}
where the isothermal speed of sound is

\begin{equation}
  c_{\rm s} = \sqrt{\frac{p_0}{\rho_0}}.
\end{equation}
The amplitude of the density oscillation $\Delta \rho= \rho - \rho_0$ is related to that
of the longitudinal velocity, $v_z$, through
\begin{equation}
  v_z = \frac{\Delta \rho}{\rho} c_{\rm s}
\end{equation}
(e.g. Landau \& Lifshitz \cite{landau}).

\begin{table}
\caption{Simulations of Alfv\'{e}n waves in a  
homogeneous medium.
The wave is characterised by the two quantities $\eta$, 
the amplitude of the imposed Alfv\'{e}n wave in terms of the vertical
magnetic field, and $\beta_{\rm }={2 \mu_0 p}/{B_z^2}$ the plasma beta.
For the different Runs we further specify the length of the time step
$\Delta t$ in terms of the period of the wave, $P$,
the length of the computational domain, $L$ in terms of the wave length, 
$\lambda$, and the number of grid points, $N$.}

\begin{tabular}{llrlll} \hline
Run & $\Delta t/P$ & $N$ & $L/\lambda$ & $\eta$   & $\beta$\\ \hline

1a & 0.003\,3 & 3\,600 & 78 & 0.007& 0.042\\
1b & 0.003\,3 & 3\,600 & 78 & 0.07 & 0.042\\
1c & 0.003\,3 & 3\,600 & 78 & 0.7 & 0.042\\

2a & 0.003\,3 & 13\,500 & 193  &  0.007 & 0.96\\
2b & 0.003\,3 & 13\,500 & 193  &  0.07  & 0.96\\
2c & 0.003\,3 & 13\,500 & 193  &  0.7 & 0.96\\ 

3a & 0.003\,3 & 13\,500 & 278 &  0.007 & 2\\
3b & 0.003\,3 & 13\,500 & 278 &  0.007  & 2\\
3c & 0.003\,3 & 13\,500 & 278 &  0.07  & 2\\

4a & 0.001\,7 & 36\,000 & 77.2 &  0.007 & 17\\
4b & 0.001\,7 & 36\,000 & 77.2 &  0.07  & 17\\
4c & 0.001\,7 & 36\,000 & 77.2 &  0.7   & 17\\ 
\hline

\label{models}
\end{tabular}
\end{table}

\subsection{Parametric decay}

In the presence of a density 
fluctuation a circularly polarised Alfv\'en wave
decays into forward propagating density and magnetic waves 
in addition to a backward propagating magnetic wave.
These waves are not necessarily normal modes of the plasma.

Galeev \& Oraevskii (\cite{go}) and Sagdeev \& Galeev (\cite{sg}) showed that an
Alfv\'{e}n wave with a frequency $\omega_0 = v_A k_0$ and a wave number 
$k_0$ can decay into a backward propagating Alfv\'{e}n wave with a
frequency $\omega_-$ and  a wave number $k_-$ and a forward propagating 
acoustic wave with a frequency $\omega$ and a wave number $k$ that 
fulfill the resonance conditions
\begin{equation}
\omega_0 = \omega + \mid{\omega_-}\mid
\end{equation}
and 
\begin{equation}
k_0=k-k_-
\end{equation}
(e.g. Cramer \cite{cramer}). 

Goldstein (\cite{gold}) and Derby (\cite{derby}) derived the dispersion relation for low 
frequency waves neglecting dispersive effects

\begin{eqnarray}
(\omega^2-c_s^2 k^2)(w- v_A k)[(w+v_A k)^2 - 4 \omega_0^2 ] \nonumber \\
=\frac{\eta v_A^2 k^2}{2}[\omega^3 + \omega^2 v_A k - 3 \omega \omega_0^2 
+ \omega_0^2 v_A k]
\label{disper}
\end{eqnarray}
where $\eta = \frac{B_{\perp}}{B_z}$.
In the limit of 
$\beta \ll 1$ Eq. (\ref{disper}) yields a maximum growth rate for the instability near
$k \simeq 2 k_0$ for the density wave.  The backward propagating magnetic
wave has then $\omega_- \simeq - \omega_0$ and $k_- \simeq k_0$ resulting
in $\omega_- \simeq - v_{\rm A} k_-$.  This means that it is a natural mode
of the system, an Alfv\'en wave, and at the same time the forward
propagating density wave is an acoustic wave.  In this regime we therefore
obtain a parametric instability.  
When $\beta$ increases the daughter wave becomes less resonant with the normal mode and 
the modulational instability loses its parametric nature.

\section{Results}

We use the numerical code of Boynton \& Torkelsson (\cite{BTT}) to simulate 
circularly polarised Alfv\'{e}n waves.  This code is based on the
ETBFCT algorithm (Boris \cite{boris}), a form of flux-corrected transport
(Boris \& Book \cite{boris:book}).  The scheme is fairly inefficient 
compared to modern methods; it needs about 30 to 40 grid points per
wave length of the Alfv\'en wave, while there are schemes that require only
a third of this.
However the memory and time requirements for a one-dimensional simulation 
are sufficiently small that this is acceptable.

The waves
are driven on the lower boundary of a one-dimensional box and propagate through a 
homogeneous medium.
In all the runs the grid is sufficiently extended that the wave does 
not hit the upper boundary during the course of the simulation.
The lower boundary condition is formulated in such a way that it only allows
an outward propagating Alfv\'en wave.  This can be a problem since backward 
propagating Alfv\'en waves are generated in the simulations.  These are mostly
of low amplitude though, and it is only very late in the simulations that 
high-amplitude Alfv\'en waves reach the lower boundary, and
the physical processes that we are interested in occur
far from the boundary then.

We study waves of different amplitudes, $\eta$, at different values of
$\beta$, the ratio of gas pressure to magnetic pressure (Tab. \ref{models}).
The magnetic field is strong in runs 1a-c, that is $v_{\rm A} > c_{\rm s}$,
intermediate in runs 2a-c and 3a-c, that is $v_{\rm A} \sim c_{\rm s}$ and 
weak in runs 4a-c,
that is $c_{\rm s} > v_{\rm A}$.

\begin{figure}
\includegraphics[width=8.8cm]{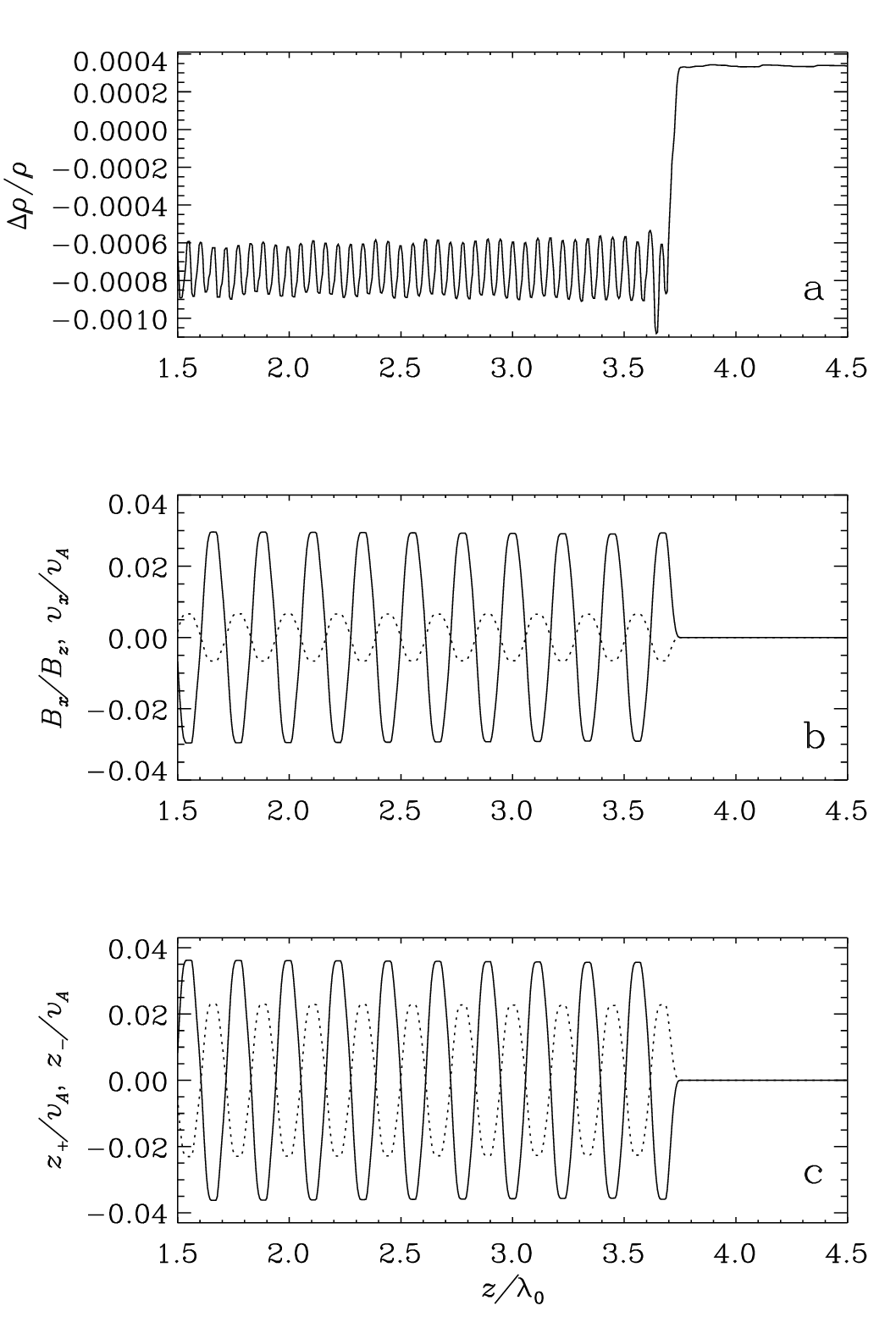}
\caption{Magnetohydrodynamic waves propagating through a homogeneous
medium of  $\beta = 0.96$ and $\eta =0.007$ (Run 2a).
{\bf a} $ \Delta \rho / \rho=\frac{ \rho - \rho_0}{\rho_0}$
versus $z/\lambda_0$ at $t/P= 16.7 $.
{\bf b} $B_x/B_z$ (solid line) and $v_x/v_A$ (dotted line) versus $z/\lambda_0$ at the same time. 
{\bf c} $z_+/v_A$ (solid line) , $z_-/v_A$ (dotted line) as 
versus $z/\lambda_0$  at the same time.}  
\label{2a}
\end{figure}


\begin{figure*}
\centering
\includegraphics[width=18cm]{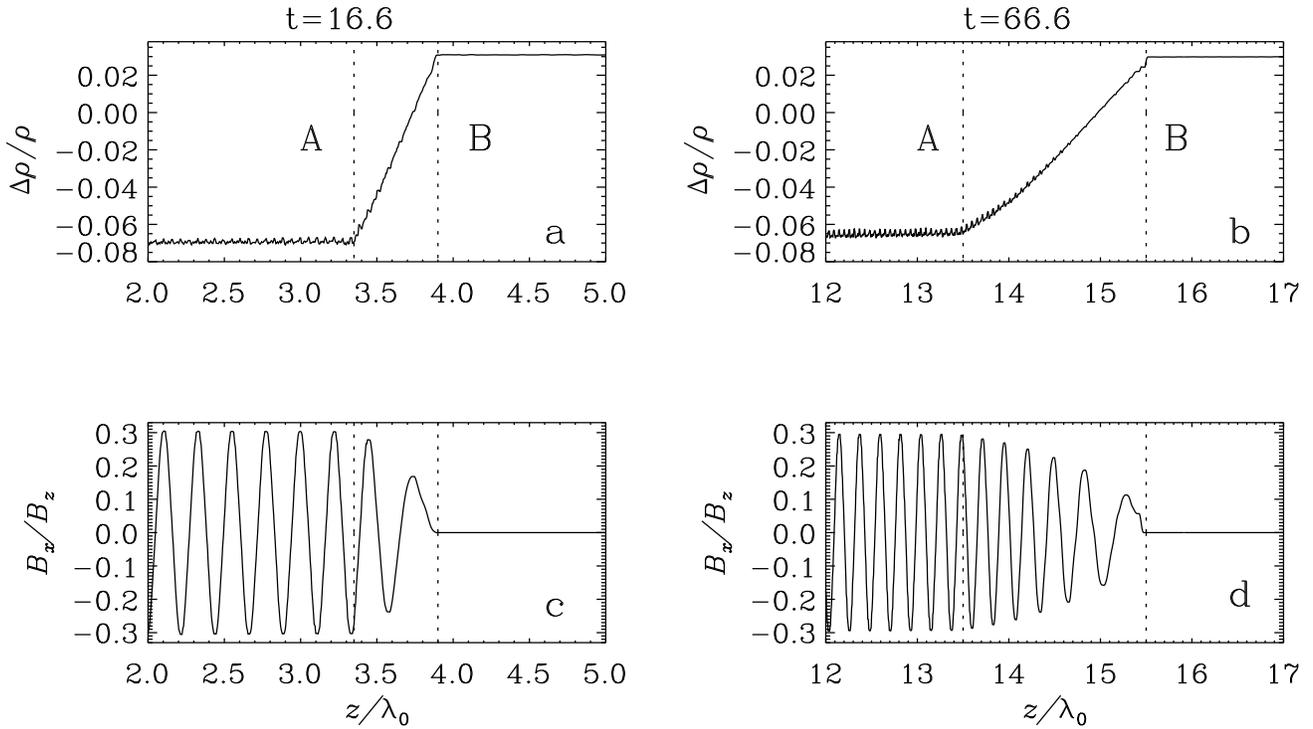}
\caption{Magnetohydrodynamic waves propagating through a homogeneous
medium of  $\beta = 0.96$ and $\eta =0.07$ (Run 2b). The plots focus
on the Alfv\'en wave front at two times, $16.6P$ ({\bf a, c}) and $66.6 P$
({\bf b, d}).  The upper panels {\bf a} and {\bf b} show $\Delta \rho/\rho$
versus $z/\lambda_0$, while the lower panels {\bf c} and {\bf d}
show $B_x/B_z$ versus $z/\lambda_0$. }
\label{1}
\end{figure*}
\subsection{Very low $\beta$} 
 
As an example we look at run 1a ($\beta = 0.042$) with a strong magnetic field and a 
low-amplitude  Alfv\'en wave ($\eta=0.007$).  
The propagating wave is shown in Fig. \ref{ph3a}b.
The Alfv\'en wave excites a density enhancement (Fig. \ref{ph3a}a), whose
right edge coincides with the front of the Alfv\'{e}n wave, while the left 
edge propagates with the lower speed $c_{\rm s}$. 
This density enhancement is a second order effect, and its amplitude is
consequently proportional to the square of the amplitude of the Alfv\'en 
wave.
One should note here that for a linearly polarised wave the corresponding 
enhancement shows a strong sinusoidal modulation (\cite{BTT}).

The density discontinuity at the front of the Alfv\'en wave excites a
secondary compressional wave which becomes more pronounced for the higher 
amplitude
 run 1b ($\eta =0.07$). This wave can be seen as a weak modulation of the
density enhancement
(Figs. \ref{ph3a}a, \ref{1b}a), and also of the magnetic pressure 
$|{\bf B_\perp}|^2/(2\mu_0)$.  The density peaks coincide with the peaks of the 
magnetic pressure making them a third order analogue of the fluctuations 
found by Boynton \& Torkelsson (\cite{BTT}, Fig. 2a).

The density fluctuations serve as the necessary seeds
of the parametric instability, but due to the low amplitude of the
fluctuations the instability grows slowly.
The evolution of the backward-propagating Alfv\'en wave that is generated by
the instability can be followed in Figs. \ref{1b}b, d, f and h.
It grows in amplitude away from the wave front,
which
enhances the growth rate of the parametric instability upstream.
Eventually, the instability is so strong that it becomes an efficient source of 
a forward-propagating sound wave (Figs. \ref{1b}a, c, e and g).
We get a region at $z < 15 \lambda_0$ in Fig. \ref{1b}g and h, in which the waves
are strongly interacting.  In this region there is a strong damping of the
forward-propagating Alfv\'en wave, and the acoustic wave is amplified until
it becomes so nonlinear that it steepens into shocks.
On the other hand since the Alfv\'en speed is larger than the sound speed, 
the head of the Alfv\'en wave manages to stay ahead of the 
density fluctuations and therefore remains essentially unaffected by the
parametric decay. 

In agreement with the theoretical prediction the sound wave has a wave number 
$k \simeq 2 k_0$.
Figures \ref{1b}d and f show that there is a phase shift of  $\frac{\pi}{2}$ 
between the backward- and the forward-propagating Alfv\'en waves as we expect 
if the backward-propagating wave is generated through the parametric 
instability.

\subsection{$\beta= 0.96$ and $2$}


We look first at the low amplitude wave 2a.  
The Alfv\'en wave is reflected against a density enhancement 
at the front of the Alfv\'en wave (Fig. \ref{2a}a and c). 
The density enhancement acts like a
wall that is pushed to the right by the wave.
Fig. \ref{2a}c shows that the reflected wave, represented by $z_-$, has
a phase shift 
of $\pi$ radians relative to the mother wave, which is characteristic of 
reflection at a fixed end.
To the left of the density enhancement (Fig. \ref{2a}a) we see rapid 
fluctuations in the 
density (cf. the low $\beta$ models).

The right edge of the density enhancement travels at a speed $\simeq c_s + v_z$
as expected from an acoustic wave, while 
the left edge of the density enhancement, that is the front of the Alfv\'en
wave, travels at $v_{\rm f} = 0.23 \, v_{\rm A} = v_{\rm A}/n$.
As a consequence of this, the measured wave length of the Alfv\'en wave is
$\lambda_0/n = 0.23 \, \lambda_0$, where $\lambda_0 = v_A \, P$.
We can see also that the wave does not show the ordinary
relation between $v_x$ and $B_x$ that we expect for an Alfv\'en wave 
(Fig. \ref{2a}b). 
While the transverse velocity component maintain the relation 
 $v_x/v_A = \, \eta$ we find for the fluctuating magnetic field
$B_x/B_z = 4.25 \, \eta = n \eta$.
This can be easily understood.  During
half a period the injected flux of the $x$-component of the magnetic field is
$\eta v_{\rm A} B_z P/\pi$, but because the wave propagates at the speed
$v_{\rm A}/n$, this magnetic flux is compressed into a distance $v_{\rm A}P/n$,
which leads to that the magnetic field is amplified by the factor $n$.

\begin{figure}
\includegraphics[width=9cm]{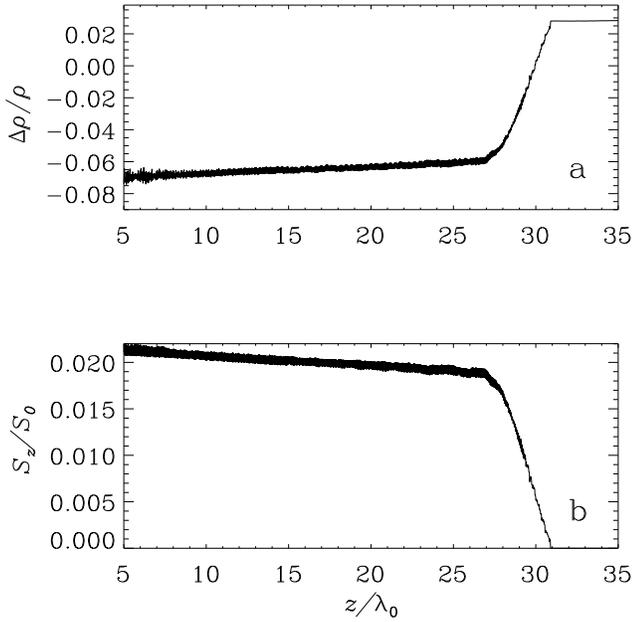}
\caption{Magnetohydrodynamic waves with $\eta = 0.07$ propagating through a homogeneous
medium of  $\beta = 0.96$ (Run 2b).
{\bf a} $ \Delta \rho / \rho=\frac{ \rho - \rho_0}{\rho_0}$
versus $z/\lambda_0$ at $t/P= 133 $.  
{\bf b} Poynting flux, $S_z$, versus $z/\lambda_0$ at the same time. $S_0$ is defined 
as $v_{\rm A} \, {B_z}^2 / \mu_0$ }
\label{2b1}
\end{figure}

The dynamics of the reflection zone is more easily studied in a wave
with a higher amplitude, because the extent of the reflection zone 
increases with the amplitude of the wave.  
We show the reflection zone of model 2b at two different times in Fig. \ref{1}.  
The density changes smoothly along a positive
slope between the positions A and B.
The slope becomes less steep with time since it is spreading out over a 
longer distance.
To the left of A the phase speed $v_{\rm p} = \lambda /P = 0.24\, v_A$,
which is the speed at which the wave front is propagating, but between
A and B the phase speed increases to $v_{\rm A}$.
The reason why the wave in run 2b runs slightly faster than the wave in run 2a
 is that it is gaining the speed $v_z$ 
of the medium in which it is propagating; a similar effect was seen in Fig. 13 
of Torkelsson \& Boynton (1998).

There is a gradual damping of the Alfv\'en wave to the left of A 
(Fig. \ref{2b1}).
This decay is accompanied by a gradual increase of the density.
These changes are negligible in the low amplitude runs 2a and 3a.
 
\begin{figure}
\includegraphics[width=9cm]{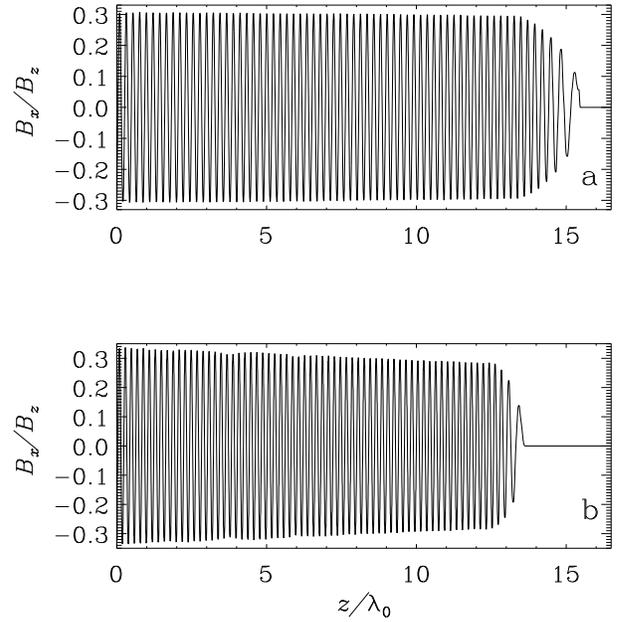}
\caption{Magnetohydrodynamic waves with $\eta = 0.07$ propagating through a homogeneous
medium of {\bf a} $\beta = 0.96$ (Run 2b) and {\bf b} $\beta = 2$ (Run 3b)
versus $z/\lambda_0$ at $t/P= 66.7$.}  
\label{23b}
\end{figure}

The higher $\beta$ models 3a and 3b share the characteristics of models 2a and 
2b (with $n= 5$) , but the damping becomes more pronounced. 
Because of the increase in the factor $n$, the wave in model 3b is  slower than the 
wave in run 2b (Fig. \ref{23b}). 
As shown in Fig. \ref{2b1}b run 2b has lost $\simeq 13.3 \%$ of its Poynting 
flux between $z = 0$ and point A at $t = 133 P$.
At the same time run 3b has lost $\simeq 26.5 \%$ of its Poynting flux.
The stronger damping is associated with a significant modulation of the Alfv\'en wave
on long length scales (Fig. \ref{23b})


\subsection{$\beta = 17$}

\begin{figure}
\includegraphics[width=8.8cm]{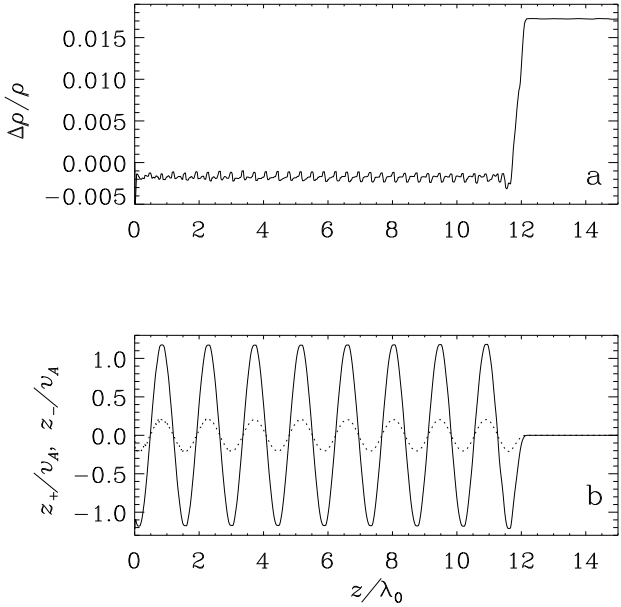}
\caption {Magnetohydrodynamic waves with $\eta = 0.7$ propagating through a homogeneous
medium of  $\beta = 17$ (Run 4c).
{\bf a} $ \Delta \rho / \rho=\frac{ \rho - \rho_0}{\rho_0}$ versus
 $z/\lambda_0$ at $t/P= 33 $.
{\bf b} $z_+/v_A$ (solid line), $z_-/v_A$ (dotted line) and
$v_x/v_A$ (dashed line) versus $z/\lambda_0$  at the same time.}
\label{f1c}
\end{figure}

\begin{figure}
\includegraphics[width=8.8cm]{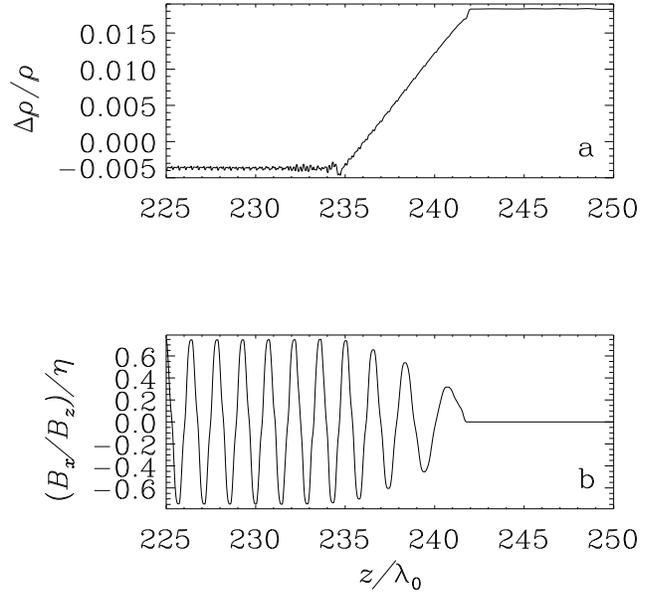}
\caption {Magnetohydrodynamic waves with $\eta = 0.7$ propagating through a homogeneous
medium of  $\beta = 17$ (Run 4c).
{\bf a} $ \Delta \rho / \rho=\frac{ \rho - \rho_0}{\rho_0}$ versus
 $z/\lambda_0$ at $t/P= 169 $.
{\bf b} $(B_x/B_z)/\eta$ versus $z/\lambda_0$  at the same time.}
\label{f2c}
\end{figure}

At a low magnetic pressure (runs 4a-c) the behaviour of the wave is almost independent of 
its amplitude. 
As an example we take the high amplitude model 4c  with $\eta=0.7$. 
The speed of the front of the Alfv\'en wave is $ \simeq 1.41 v_A$ in models
4a and 4b and $\simeq 1.43 \, v_A$ in model 4c. 
The higher wave speed in run 4c is the result of that the Alfv\'en wave has
generated an outflow with a speed $\sim 0.02 \, v_{\rm A}$.
Analogously to models 2 and 3, the amplitude of the magnetic field has 
changed with a factor $0.7 = \frac{1}{1.43}$ to compensate for the increase in
the speed of the wave front (Fig. \ref{f2c}b). 
%
The change from sub-Alfv\'enic waves in runs 2 and 3 to super-Alfv\'enic waves
in runs 4 is accompanied by a change in the character of the reflection.
Fig. \ref{f1c} shows that there is no phase delay between the two Els\"{a}sser 
variables $z_\pm$, which is a characteristic of the reflection of a wave at a free end.

Fig. \ref{f2c} shows the behavior of the wave close to the reflection zone. 
As in models 2 and 3, the slope in density  
becomes less steep with time (compare Fig. \ref{f1c}a and Fig. \ref{f2c}a).
The behaviour of the waves within this zone, in terms of the amplitude and the phase 
speed is similar to that described in sect. 3.2 for models 2 and 3 (with $n = 1.43$). 
The difference is that  the damping of the Alfv\'en wave outside the 
reflection zone is too small to be measured.

\section{Discussion}

In constructing models of Alfv\'en wave driven stellar winds, it is
important to understand the mechanism by which the Alfv\'en waves are
damped.  A linearly polarised Alfv\'en wave of high amplitude can
steepen and form current sheets even in a homogeneous medium, which
leads to a quick damping of the wave.
While circularly polarised Alfv\'en waves cannot steepen in this way,
they can be subject to a parametric decay into a backward propagating magnetic
wave and a forward propagating density wave.
Our simulations show 
that the compression of the background medium that takes place at
the wave front of an Alfv\'en wave in a low $\beta$ plasma is sufficient
to trigger this instability.  At $\beta \ga 1$ this is not sufficient to
trigger the parametric decay, but instead we find that the Alfv\'en wave 
is reflected at the wave front, whose speed deviates from the Alfv\'en
speed.

One can find a large range of $\beta$s in the solar corona and solar wind.
Typically $\beta \ll 1$ in the lower solar corona, and $\beta \ga 1$ in
the solar wind (Gary \cite{gary}).  This means that different processes
may affect Alfv\'en waves in different parts of the solar corona/wind.
The parametric decay should be at work in the lower solar corona, and
in the presence of sufficiently large fluctuations it may also be at 
work in the solar wind as indicated by the numerical simulations of
Del Zanna et al. (\cite{zanna}).  Therefore either the parametric decay
or reflection at the wave front may be the source of the inward propagating
Alfv\'en waves that have been detected in the solar corona (e.g. 
Bavassano et al. \cite{BV}).   

An interesting property of the parametric decay is that it is relatively
weak at the wave front, but the backward propagating
Alfv\'en wave that is generated serves as a seed for the instability
further upstream, where the growth rate therefore becomes larger.  
The area close to
the wave front is therefore not significantly affected by the parametric
decay, though the decay becomes strong some distance behind the wave front.
This gives rise to a self-limitation of the length of an Alfv\'en wave
packet, and it also suggest that, at least in a low-$\beta$ plasma,
the Alfv\'en wave is the least turbulent right behind the wave front.
In a high-$\beta$ plasma on the other hand one may expect to see a gradual
change in density right behind the wave front, and further upstream only
oscillations around a constant density.

\section{Conclusions}

In this paper we have presented numerical simulations of Alfv\'en waves
in homogeneous media with different
magnetic field strengths.
In a strongly magnetised plasma the Alfv\'en wave decays parametrically to
a backward propagating Alfv\'en wave and a forward propagating acoustic wave
unless the 
amplitude of the Alfv\'en wave is very low.  However by its very nature the 
parametric decay is fairly inefficient at the wave front, even when it is 
the cause of a strong damping upstream.  

For $\beta \sim 1$ or higher
we find a different pattern.
The wave front is propagating at the speed $v_{\rm A}/n$ and the
magnetic field fluctuations are amplified by a factor $n$.  $n > 1$ 
for $\beta \sim 1$ but 
$n < 1$ for $\beta \gg 1$. 
At the wave front the Alfv\'en wave is reflected by an extended region with 
a positive density gradient.  
The spatial extent of this region increases with time.
Inside this region the phase speed of the Alfv\'en wave is 
increasing from $v_{\rm A}/n$ to $v_{\rm A}$.
When $n > 1$ the reflected Alfv\'en wave is phase-shifted by $\pi$, while
it does not suffer any phase-shift for $n < 1$.  There is also a 
significant gradual damping, $\sim 10 - 20 \%$ of the Alfv\'en wave when $n > 1$.


The results presented in this paper are derived from  a highly simplified 
model, which allows us to isolate a few physical effects and study them in
detail.
This study can guide us in future investigations of the dynamics of Alfv\'en waves 
in more realistic configurations. 
In forthcoming papers  we will discuss the effects of stratification and an
expanding magnetic field.


\begin{acknowledgements}
This research was sponsored by the Swedish Research Council.  RT wants to
thank the solar MHD group at St. Andrews for hospitality during a part of this
work. UT and RT thank UKAFF, University of Leicester for hospitality during the completion of this paper. The visitor's programme at UKAFF is sponsored by the  EU Fifth Framework Programme.

\end{acknowledgements}

\end{document}